\documentstyle[12pt,epsfig]{article}

\def\plb#1{Phys.~Lett.~{\bf B#1}}
\def\npb#1{Nucl.~Phys.~{\bf B#1}}
\def\prl#1{Phys.~Rev.~Lett.~{\bf #1}}
\def\prd#1{Phys.~Rev.~{\bf D#1}}

\def\beq{\begin{equation}}
\def\eeq{\end{equation}}
\def\lagrange{{\cal{L}}}

\begin{document}

\begin{titlepage}
 \hfill       OHSTPY-HEP-T-01-023\\
  \mbox{ } \hfill      August, 2001 \\

\vspace{1.0cm}

\begin{center}

  {\Large\bf SO(10) grand unification in five dimensions:}

\vspace{.5cm}

  {\Large\bf Proton decay and the $\mu$ problem}

  \vspace{2cm}

  {\large\bf Radovan Derm\' \i \v sek$^1$ and Arash Mafi$^{1,2}$}

    \bigskip
{\em $^1$Department of Physics, The Ohio State University, \\
174 W. 18th Ave., Columbus, Ohio  43210\\
$^{2}$Department of Physics, University of Arizona, \\
 Tucson, Arizona  85721}

  \vspace{1cm}
{\bf Abstract}

\end{center}

\noindent We construct a minimal supersymmetric $SO(10)$ grand
unified model in 5 dimensions. The extra dimension is compactified
on an $S^1/(Z_2 \times Z_2^\prime)$ orbifold which has two
in-equivalent fixed points. These are flat 4-dimensional Minkowski
spaces: the visible and the hidden branes. By orbifolding, the gauge
symmetry on the hidden brane is reduced down to the Pati--Salam gauge
symmetry $SU(4) \times SU(2)_L \times SU(2)_R$. On the visible
brane the $SO(10)$ is broken by the ordinary Higgs mechanism down to
$SU(5)$. The resulting 4-dimensional theory has the standard model
gauge symmetry (the intersection of $SU(5)$ and $SU(4) \times
SU(2)_L \times SU(2)_R$) and the massless spectrum consists of
the MSSM gauge fields and two Higgs doublets. The matter fields are
assumed to live on the visible brane. We discuss gauge coupling
unification in our 5-dimensional model in terms of corrections to
the conventional 4-dimensional unification. Supersymmetry is
broken on the hidden brane (where mass terms for gauginos and a
$\mu$ term are generated) and communicated to squarks and sleptons
via gaugino mediation. We also discuss a possibility of linking
the supersymmetry
breaking on the hidden brane to
the Higgs mechanism responsible for partial breaking of the gauge
symmetry on the
visible brane via the shining mechanism.
Finally, there are no operators of
dimension 5 leading to proton decay.
Proton decay through dimension 6 operators is enhanced
compared to
conventional GUTs and can be seen in current or next generation proton
decay experiments.

\end{titlepage}

\section{Introduction}
\label{sec:intro}
The quest for a unified picture of particles and gauge interactions
has led physicists to consider the grand unified theories (GUTs) as
serious candidates for physics at high energies. 
GUTs offer a simple explanation of the quark and lepton quantum
numbers \cite{su4xsu2xsu2, su5, so10} and the minimal
supersymmetric framework
leads to a successful prediction of the weak mixing angle from gauge
coupling unification at the scale $M_G\approx 10^{16}$ GeV \cite{susygut,
gutexp}.
Beyond the grand unification scale, one expects to see effects coming
from the Planck scale physics which are hoped to be explained in the
context of superstring theory. In order to get the 4-dimensional
space-time from a 10-dimensional string theory, the extra
dimensions need to be compactified. Therefore the
ideas of supersymmetry, grand unification and extra dimensions
are direct consequences of our quest for a unified picture of physics.

At the string scale, the specific compactification
dynamics chosen by nature
leads to the particular pattern of particles and
symmetries. The nature of
this dynamics is yet to be understood. We tend to follow a bottom up
approach. We try to speculate about the high energy
dynamics based on the low energy physics
that has the advantage of being examined by experiments. 
The compatibility of the consistent theories at high energies with the
low energy measurments is a non-trivial test for the candidates
that are hoped to come out of the string theory.

The search for alternatives among theories constructed in higher 
dimensional spacetimes is motivated by the fact that conventional 
SUSY GUTs face many problems that remain to be answered in order
to give a more complete picture of nature.
Some problems like the proton decay push the conventional SUSY GUT 
models to the edges of viability \cite{pdecay} and cast 
a shadow of doubt on our understanding of nature beyond the 
electroweak (EW) scale. A few important other questions include
the nature of SUSY breaking and mediation, the $\mu$ problem, 
suppression of flavor changing neutral current 
effects,
GUT breaking mechanism and the doublet triplet splitting problem.

The recent interest in this direction at the field theory level
started after the work of Kawamura \cite{Kawamura} which provides 
an elegant way of an $SU(5)$ symmetry breaking and doublet triplet 
splitting by an orbifold compactification of a theory
formulated in 5-dimensions.\footnote{We note that using 
orbifolds to reduce a gauge 
symmetry was introduced in string phenomenology \cite{DHVW}.} 
The framework was further developed in
\cite{Altarelli+Feruglio, Hall+Nomura} and nice examples of $SU(5)$
models in 5-dimensions with fifth dimension compactified on the $S^1/(Z_2
\times Z_2^\prime)$ orbifold were constructed \cite{Altarelli+Feruglio,
Hall+Nomura, Hebecker+March-Russel_l}. We find it interesting to see what
can be achieved by an orbifold compactification in the case of $SO(10)$
gauge symmetry in 5-dimensions. 

In this paper, we assume that the physics at some high energy
scale can be described by a 5-dimensional $SO(10)$ SUSY GUT. 
We assume the lowest amount of supersymmetry with a minimal particle
content in the 5-dimensional bulk. 
The extra dimension is compactified
on a $S^1/(Z_2 \times Z_2^\prime)$ orbifold which has two
in-equivalent fixed points. These are flat 4-dimensional Minkowski
spaces: the visible and the hidden branes. By orbifolding, the gauge
symmetry on the hidden brane is reduced down to the Pati--Salam gauge
symmetry $SU(4) \times SU(2)_L \times SU(2)_R$. On the visible
brane the $SO(10)$ is broken by ordinary Higgs mechanism down to
$SU(5)$. The resulting 4-dimensional theory has the standard model
gauge symmetry (the intersection of $SU(5)$ and $SU(4) \times
SU(2)_L \times SU(2)_R$) and the massless spectrum consists of
MSSM gauge fields and two Higgs doublets. The matter fields are
assumed to live on the visible brane. The model is described 
in section \ref{sec:SO(10)model}. In section \ref{sec:unification} we
discuss gauge coupling
unification in our 5-dimensional model in terms of corrections to 
the conventional 4-dimensional unification comming from heavy Kaluza-Klein
modes of the gauge and Higgs fields. 

The advantage of higher dimensional construction
over the conventional SYSY GUT models is that it provides a
framework with a potential to solve several challenging questions
that the conventional GUTs face. In section \ref{sec:susybreak}
we discuss a mechanism for supersymmetry breaking. Supersymmetry is
broken on the hidden brane where mass terms for gauginos    
are generated. Its breaking is communicated to squarks and
sleptons via gaugino mediation \cite{gMSB} which explains 
the suppression of
flavor changing neutral current effects.
We also discuss a possibility of linking the SUSY breaking on the hidden brane to 
the Higgs mechanism responsible for partial breaking of the gauge symmetry on the 
visible brane via the shining mechanism of Ref. \cite{shining}.
A solution to the $\mu$ problem is 
proposed in section \ref{sec:muterm}
%and a solution to doublet-triplet splitting problem 
and suppression of proton
decay is discussed in section \ref{sec:protondecay}.
Finally we conclude in section \ref{sec:conclusions}.

During preparation of this article, works \cite{so(10)in6d} appeared
considering $SO(10)$ symmetry breaking by orbifold compactification in
6-dimensions.

\section{Minimal SUSY SO(10) model in five dimensions}
\label{sec:SO(10)model} In five dimensions, $\cal{N}$~=~1 supersymmetry
is generated by 8 supercharges and is equivalent to $\cal{N}$~=~2
supersymmetry in four dimensions. A trivial spatial
compactification of $\cal{D}$~=~5,~$\cal{N}$~=~1 results in
$\cal{N}$~=~2 supersymmetry in 4 dimensions. However, for the
purposes of model building, it is desirable to have $\cal{N}$~=~1
supersymmetry in 4 dimensions and orbifolding is an elegant way of
reducing the supersymmetry. Furthermore, orbifolding can also be
used to break the gauge symmetry in a grand unified theory
\cite{Kawamura}.

In this section we present a minimal $\cal{D}$~=~5,~$\cal{N}$~=~1
supersymmetric model with $SO(10)$ GUT gauge symmetry and
additional structure on the orbifold fixed points. We compactify
the extra dimension on an orbifold $S^1/(Z_2\times Z^\prime_2)$ to
reduce the supersymmetry from $\cal{N}$~=~2 to $\cal{N}$~=~1 and
also reduce the $SO(10)$ gauge symmetry. The complete breaking of
the $SO(10)$ gauge symmetry to the Standard Model (SM) gauge group
$SU(3)\times SU(2)\times U(1)\equiv G(SM)$ is achieved via a
combination of orbifolding and Higgs mechanism. The massless
sector corresponds to the usual spectrum of MSSM.

\subsection{The $S^1/(Z_2\times Z^\prime_2)$ orbifold}
\label{sec:orbifold} We consider a 5-dimensional space-time with
the 5'th dimension compactified on an orbifold. Following
Ref.~\cite{Kawamura,Altarelli+Feruglio,Hall+Nomura} the orbifold
is taken to be $S^1/(Z_2\times Z^\prime_2)$ where $S^1$ is a
circle of radius $R=1/M_c\sim 1/M_G$ defined with a periodic coordinate
$0\leq x_5<2\pi R$. $S^1/Z_2$ is obtained by dividing $S^1$ with a
$Z_2$ transformation $x_5 \to -x_5$. We further divide $S^1/Z^2$
orbifold by a $Z^\prime_2$ transformation $x_5^\prime\to
-x_5^\prime$ with $x_5^\prime=x_5+\pi R/2$ to obtain
$S^1/(Z_2\times Z^\prime_2)$. $x_5=0$ and $x_5=\pi R/2$ are
in-equivalent orbifold fixed points. These fixed points are each a
flat 4-dimensional Minkowski space and we refer to them as the
visible brane and the hidden brane, respectively. (Later in
section \ref{sec:matter} the ordinary matter fields are assumed to
be confined to the visible brane.) A generic field
$\phi(x_\mu,x_5)$ in the 5-dimensional bulk is identified by its
transformations under the $Z_2$ and $Z^\prime_2$ parities $P=\pm$
and $P^\prime=\pm$, respectively.
\begin{eqnarray}
\nonumber
\phi(x_\mu,x_5)\to \phi(x_\mu,-x_5)&=&P~\phi(x_\mu,x_5),\\
\phi(x_\mu,x_5^\prime) \to \phi(x_\mu,-x_5^\prime)&=&P^\prime
~\phi(x_\mu,x_5^\prime).
\end{eqnarray}
A field $\phi_{\pm\pm}(x_\mu,x_5)$ with a definite set of  parities
$(P,P^\prime)=(\pm,\pm)$ has a unique Fourier series expansion:
\begin{eqnarray}
\nonumber
\phi_{++}(x_\mu,x_5)&=&\frac{1}{\sqrt{2^{\delta_{n,0}}\pi R}}
\sum^{\infty}_{n=0}
\phi^{(2n)}_{++}(x_\mu) cos\frac{2nx_5}{R},\\
\nonumber
\phi_{+-}(x_\mu,x_5)&=&\frac{1}{\sqrt{\pi R}} \sum^{\infty}_{n=0}
\phi^{(2n+1)}_{+-}(x_\mu) cos\frac{(2n+1)x_5}{R},\\
\nonumber
\phi_{-+}(x_\mu,x_5)&=&\frac{1}{\sqrt{\pi R}} \sum^{\infty}_{n=0}
\phi^{(2n+1)}_{-+}(x_\mu) sin\frac{(2n+1)x_5}{R},\\
\phi_{--}(x_\mu,x_5)&=&\frac{1}{\sqrt{\pi R}} \sum^{\infty}_{n=0}
\phi^{(2n+2)}_{--}(x_\mu) sin\frac{(2n+2)x_5}{R}
\end{eqnarray}
From the 4-dimensional perspective
the Fourier component fields (Kaluza-Klein states)
$\phi^{(2n)}_{++}$ acquire a mass $2n/R$, $\phi^{(2n+1)}_{+-}$
and $\phi^{(2n+1)}_{-+}$ a mass $(2n+1)/R$, and
$\phi^{(2n+2)}_{--}$ a mass $(2n+2)/R$. Only $\phi_{++}(x_\mu,x_5)$
has a massless Kaluza-Klein (KK) mode $\phi^{(0)}_{++}(x_\mu)$ and all
other KK modes have mass of order GUT scale or larger.
$\phi_{++}(x_\mu,x_5)$ and $\phi_{+-}(x_\mu,x_5)$ are non-vanishing on
the visible brane and can directly couple to the ordinary matter living
on the visible brane. $\phi_{++}(x_\mu,x_5)$ and $\phi_{-+}(x_\mu,x_5)$
are non-vanishing on the hidden brane.
On the other hand, the fields $\phi_{-+}(x_\mu,x_5)$ and
$\phi_{--}(x_\mu,x_5)$ ($\phi_{+-}(x_\mu,x_5)$ and $\phi_{--}(x_\mu,x_5)$)
have non-vanishing $x_5$-derivatives on the visible (hidden) brane
and their $x_5$-derivatives can couple directly to the fields on the
visible (hidden) brane .

\subsection{Gauge symmetry structure on the orbifold }
\label{sec:SO(10)andorbifold}

$\cal{N}$~=~1 SUSY in $\cal{D}$~=~5 space-time may be formulated
in terms of the usual $\cal{D}$~=~4,~$\cal{N}$~=~1 superfield
notation. A $\cal{D}$~=~5,~$\cal{N}$~=~1 gauge
supermultiplet can be decomposed into a
$\cal{D}$~=~4,~$\cal{N}$~=~1 gauge ($V$) and a chiral ($\phi$)
supermultiplet. In the same way, a $\cal{D}$~=~5,~$\cal{N}$~=~1
hypermultiplet can be expressed as a pair of
$\cal{D}$~=~4,~$\cal{N}$~=~1 chiral multiplets. In
our model, we assume that a single $\cal{D}$~=~5,~$\cal{N}$~=~1
gauge supermultiplet and Higgs hypermultiplet live in the bulk.
The gauge supermultiplet is in the adjoint representation
(45-dimensional) of $SO(10)$. We take the Higgs hypermultiplet to be
in the 10-dimensional representation of $SO(10)$ and refer to the
$\cal{N}$~=~1 chiral superfield components as $10_H$ and
$10^\prime_H$. The action for our minimal model containing one
gauge and one massless hypermultiplet in the bulk can be expressed
in terms of the $\cal{N}$~=~1 superfields as~\cite{AGW}
\begin{eqnarray}
\label{eq:action}
\nonumber
S_{N=2}&=&\int{d^5x}\Bigg\{\frac{2}{g^2}
Tr\Bigg[\frac{1}{4}\int{d^2\theta W^\alpha W_\alpha}+h.c.
\qquad\qquad\qquad\\
\nonumber &+&\int{d^4\theta\Bigg((\sqrt{2}\partial_5+\phi^\dag})
e^{-V}(-\sqrt{2}\partial_5+\phi)e^V+
\partial_5 e^{-V}\partial_5 e^V\Bigg)\Bigg]\\
\nonumber &+&\int{d^4\theta}[10_H^\prime e^V 10_H^{\prime \dag} +
10_H^\dag e^{-V}10_H]\\
&+&\Bigg[\int{d^2\theta}10_H^\prime(\partial_5-\frac{1}{\sqrt{2}}
\phi)10_H+h.c.\Bigg]\Bigg\}.
\end{eqnarray}
The action is invariant under the $Z_2$ transformations:
\begin{eqnarray}
V(x_\mu,x_5)\to V(x_\mu,-x_5)&=&P ~V(x_\mu,x_5)~ P\\
\phi(x_\mu,x_5)\to \phi(x_\mu,-x_5)&=&-P~ \phi(x_\mu,x_5) P\\
10_H(x_\mu,x_5)\to 10_H(x_\mu,-x_5)&=&P ~10_H(x_\mu,x_5)\\
10_H^\prime(x_\mu,x_5)\to 10_H^\prime(x_\mu,-x_5)&=&-P^T
~10_H^\prime(x_\mu,x_5)
\end{eqnarray}
where $P$ is a $10\times 10$ matrix acting on the gauge indexes and
$P^2=1$. The action is
also invariant under the $Z^\prime_2$ transformations where $P^\prime$ and
$x_5^\prime=x_5+\pi R/2$ replace $P$ and $x_5$. We choose
$P=1_{5\times 5}\otimes 1_{2\times 2}$ and
$P^\prime= \, \rm diag \, (-1,-1,-1,1,1)\otimes 1_{2\times 2}$.
The $Z_2\times Z^\prime_2$ charges of the superfields are listed
in the table.
\begin{table}[h]
$$
\begin{array}{|c|c|c|}
\hline
10_H& Z_2\times Z^\prime_2& mass \\ \hline \hline
6_H& (+,-)& (2n+1)/R \\ \hline
4_H& (+,+)& 2n/R \\ \hline
\end{array}
\qquad
\begin{array}{|c|c|c|}
\hline
V& Z_2\times Z^\prime_2&mass \\ \hline \hline
V_{t_+}& (+,+)&2n/R \\ \hline
V_{t_-}& (+,-)&(2n+1)/R \\ \hline
\end{array}
$$
%\end{table}
%\vspace{-.3in}
%\begin{table}[h]
$$
\begin{array}{|c|c|c|}
\hline
10_H^\prime& Z_2\times Z^\prime_2&mass \\ \hline \hline
6_H^\prime& (-,+)&(2n+1)/R \\ \hline
4_H^\prime& (-,-)& (2n+2)/R\\ \hline
\end{array}
\qquad
\begin{array}{|c|c|c|}
\hline
\phi& Z_2\times Z^\prime_2&mass \\ \hline \hline
\phi_{t_+}& (-,-)&(2n+2)/R \\ \hline
\phi_{t_-}& (-,+)&(2n+1)/R \\ \hline
\end{array}
$$
\caption{The decomposition of 4d vector supermultiplet V, chiral
supermultiplet $\phi$, and chiral multiplets $10_H$ and
$10^\prime_H$ according to their parity assignments with
corresponding KK masses.}
\end{table}

$10_H$ and $V$ have even $Z_2$ parities, while $10_H^\prime$ and
$\phi$ have odd $Z_2$ parities and vanish at $x_5=0$. This signals
the breakdown of $\cal{N}$~=~2 supersymmetry into $\cal{N}$~=~1.
The non-vanishing $10_H$ and $V$ on the visible brane are complete
$SO(10)$ multiplets and the $SO(10)$ gauge symmetry is respected
on the visible brane.

The $Z_2^\prime$ projection breaks the $SO(10)$ gauge group into
$SO(6)\times SO(4)$ on the hidden brane. The three $-1$'s in
$P^\prime$ are associated with the $SO(6)$ and the two $+1$'s are
related to the $SO(4)$. These and following observations are
elaborated in detail in the appendix. Note, that $SO(6)\sim SU(4)$
and $SO(4)\sim SU(2)_L\times SU(2)_R$. In fact, $SU(4)$ contains
$SU(3)\times U(1)_{B-L}$ where $SU(3)$ is the SM QCD gauge group
and $U(1)_{B-L}$ is the symmetry group associated with the baryon
number minus lepton number generator. $SU(2)_L$ is the weak gauge
group.

The Higgs fields $10_H$ and $10_H^\prime$ are realized under
$SO(6)\times SO(4)$ subgroup of $SO(10)$ as $10=6\oplus 4$.  Under
the SM gauge group, we further have $6=t\oplus \bar{t}$ and
$4=d\oplus \bar{d}$ where $t~(\bar{t})$ is a color
triplet~(anti-triplet) and $d, \, \bar{d}$ are weak doublets.

Based on the notation used in the appendix, $V$ and $\phi$ (which
are in the adjoint representation of $SO(10)$) are classified into
$t_+$-type and $t_-$-type categories. The $t_+$-types belong to
the $SO(6)\times SO(4)$ subgroup of $SO(10)$ and commute with
$P^\prime$. The $t_-$-types belong to $SO(10)/(SO(6)\times SO(4))$
and anti-commute with $P^\prime$.

From the parity assignments in the table we see that only the two
Higgs doublets, $d_H$ and $\bar{d}_H$ (contained in $4_H$), and
the gauge fields $V_{t_+}$ of $SO(6)\times SO(4)$ gauge group have
zero modes.

To summarize the orbifolding results, at the zero-mode level, we
are left with $\cal{N}$~=~1  supersymmetry and $SO(6)\times SO(4)$
gauge symmetry. $SO(10)$ gauge symmetry is respected on the
visible brane while there is only $SO(6)\times SO(4)$ gauge
symmetry on the hidden brane. There is $\cal{N}$~=~1 supersymmetry
on both branes.

We now need to further reduce the $SO(6)\times SO(4)$ gauge
symmetry of the zero-modes to the SM gauge symmetry.
Unfortunately, this is not possible to achieve by a $Z_2$
projection. In general is in not possible to break the rank of a
group by an abelian orbifolding symmetry. This applies to the case
of the orbifold breaking by inner automorphism. In the case of the
orbifold breaking by outer automorphisms the rank reduction is
possible but only in very limited ways. The $SO(10)$ symmetry
cannot be reduced in this way to $SU(5)$ or $G(SM)$. These issues
were recently discussed in detail in Ref.
\cite{Hebecker+March-Russell_2}.

The reduction of the $SO(6)\times SO(4)$ gauge symmetry to the SM
gauge group can be accomplished via the ordinary Higgs mechanism
on either the visible or the hidden brane. We assume that a pair
of $16, \bar{16}$ living on the visible brane gets a vacuum
expectation value of order $M_c$ in the right handed neutrino
direction.\footnote{Another possibility is to assume that a field
transforming as $(4,2)$ under $SO(6)\times SO(4)$ lives on the
hidden brane and gets a vev that breaks the gauge symmetry down to
the SM guage group.} This breaks $SO(10)$ down to $SU(5)$ on the
visible brane and gives GUT-scale mass to those gauge fields among
$V_{t_+}$ that belong to $SO(6)\times SO(4)/G(SM)$ including the
orbifold zero-modes. Since $SO(10)$ is broken on the hidden brane
to $SO(6)\times SO(4)$ we are left with the SM gauge symmetry in
the four dimensional theory (the intersection of $SU(5)$ and
$SO(6)\times SO(4)$). The massless spectrum consists of MSSM gauge
fields and two Higgs doublets.

\subsection{MSSM matter fields}
\label{sec:matter} Our next task is to identify the ordinary
matter and other fields living on the branes along with all
possible interactions of the brane and bulk fields. The
interactions should respect the $SO(10)$ gauge symmetry on the
visible brane and $SO(6)\times SO(4)$ on the hidden brane.

There are several approaches explored in Refs.
\cite{Altarelli+Feruglio, Hall+Nomura,Hebecker+March-Russel_l}
within the context of SU(5). Realistic fermion
masses can be obtained by mixing brane--localized multiplets with
bulk multiplets \cite{Hall+Nomura}, or it is possible to have a
complete freedom in Yukawa couplings if matter fields originate
from different bulk multiplets \cite{Hebecker+March-Russel_l}.
Many of these mechanisms can be generalized to the SO(10) case.
Since fermion masses are not the subject of this paper we assume
here the simplest possibility that ordinary matter superfields are
localized on the visible brane. Each family of ordinary matter
fermions and their superpartners is represented by a chiral
superfield in the 16-dimensional representation of $SO(10)$. For the
purposes of this paper, we only consider the third generation and
refer to it as $16_3$.

In order to construct an action for the matter fields on the
visible brane, we need to know their $Z_2\times Z_2^\prime$
transformation properties. After all, we are interested in having
an action which is invariant under the $Z_2\times Z_2^\prime$
symmetry. The $Z_2$ parity of the matter fields living on the
visible brane must all be plus. The $Z_2^\prime$ parity of these
fields are determined by requiring that any $SO(10)$ invariant
operator on the visible brane transforms covariantly under
$Z_2^\prime$. $Z_2^\prime$ orbifold identifies the points $x_5=0$
and $x_5=\pi R$. If various parts of an $SO(10)$ invariant
operator at $x_5=0$ transform differently under $Z_2^\prime$, the
corresponding terms at $x_5=\pi R$ cannot combine into an $SO(10)$
invariant operator. To our choice of $P^\prime$ acting on $10$
dimensional representation correspond these possible $P^\prime$
charge assignments of fields in a spinor representation
$P^\prime(Q,U,D,L,E,\nu)=\pm(+,-,-,+,-,-)$. The invariant action
for $16_3$ coupled to the Higgs superfield non-vanishing on the
visible brane ($10_H$) is 
\beq \label{eq:mattera}
S_{matter}=\int{d^5x} \, \frac{1}{2}
\left\{\delta(x_5)-\delta(x_5-\pi R)\right\} \sqrt{2\pi R}
\lambda_3 \, \int{d^2\theta} \; 16_3 10_H 16_3 + {\rm h.c.}, 
\eeq
where $\lambda_3$ is the dimensionless Yukawa coupling. The
coupling of the ordinary matter supermultiplet $16_3$ to the Higgs
$10_H$ generates the desired MSSM couplings of the quark and
lepton superfields to the massless Higgs zero-mode doublets.
Integrating out the extra dimension and writing the effective
4-dimensional Lagrangian in terms of the KK modes of the bulk
fields we end up with
\begin{eqnarray}
\label{eq:matterb} \nonumber
\lagrange_4&=&\sqrt{2}\lambda_3\sum^{\infty}_{n=0}\int{d^2\theta}\big[
\frac{1}{\sqrt{2^{\delta_{n,0}}}}Q U d^{(2n)}_H+
\frac{1}{\sqrt{2^{\delta_{n,0}}}}Q D \bar{d}^{(2n)}_H+
\frac{1}{\sqrt{2^{\delta_{n,0}}}}L \nu d^{(2n)}_H  \\
\nonumber &+&\frac{1}{\sqrt{2^{\delta_{n,0}}}}L E
\bar{d}^{(2n)}_H+ \frac{1}{2} Q Q t^{(2n+1)}_H+ U E t^{(2n+1)}_H+
Q L \bar{t}^{(2n+1)}_H+ U D \bar{t}^{(2n+1)}_H \big] + {\rm h.c.}.\\
\end{eqnarray}
The Higgs zero-modes are coupled exactly in the form of the
MSSM.
However, a $\mu$ term is absent in the superpotential and the
appropriate mechanism to generate the $\mu$ term at the right
scale will be presented in section \ref{sec:muterm}.

$10_H^\prime$ vanishes on the visible brane and therefore there is
no coupling in the superpotential like $16_3 10_H^\prime 16_3$.
However there are other possible interactions on the branes that
do not vanish either on the visible or hidden brane. Consider an
interaction on the visible brane 
\beq 
\label{eq:forbida}
\int{d^5x}\frac{1}{2}\{\delta(x_5)+\delta(x_5-\pi
R)\} \int{d^2\theta} \, 10_H 10_H. 
\eeq 
Similar interactions
respecting the $SO(6)\times SO(4)$ symmetry can in principal be
present on the hidden brane too, 
\beq \label{eq:forbidb}
\int{d^5x}\frac{1}{2}\{\delta(x_5-\pi
R/2)+\delta(x_5+\pi R/2)\} \int{d^2\theta} \, (\ \lambda_6
6^\prime_H 6^\prime_H \ + \ \lambda_4 4_H 4_H \ ). 
\eeq 
The Higgs triplets are coupled in such terms by a mass of order $M_c$
and seem to confront our model with the usual challenge of the conventional
SUSY GUT models -- proton decay.
Another major complication expected from these terms is that they couple the
zero-mode Higgs doublets $d^{(0)}_H$ and $\bar{d}^{(0)}_H$ in the
superpotential and produce a GUT scale $\mu$ parameter that is
too large. To cure this problem, we notice that the action
(\ref{eq:action}) is invariant under a $U(1)_R$ symmetry and also a
vector-like $U(1)_{PQ}$ Peccei-Quinn (PQ) symmetry under which
$10_H$ and $10_H^\prime$ have opposite charges. We see that
interactions in Eqs. (\ref{eq:forbida}), (\ref{eq:forbidb}) can be
easily forbidden by requiring that any of these symmetries is
respected on branes. Later in section \ref{sec:muterm} we will
find convenient to assume that the full theory is invariant under
$U(1)_{PQ}$ with $Q_{10_H} = +1$ and $Q_{10^\prime_H} = -1$.
\section{Gauge coupling unification}
\label{sec:unification} In this chapter, we discuss the issue of
gauge coupling unification. The massless spectrum in this model is
exactly that of the MSSM. The gauge couplings run from the 
EW scale to the
compactification scale $(M_c\sim 1/R)$ with the ordinary MSSM
$\beta$ functions. They almost unify at $M_c$ which is below but
very close to the conventional GUT scale. Beyond $M_c$, the gauge
couplings run slowly due to the heavy KK modes that do not fill
degenerate GUT multiplets and unify at scale $M_*$. Such threshold
corrections coming from the KK states can be easily estimated
\cite{DDG,Hall+Nomura}. If we assume a unified value $\alpha_*$
at a scale $M_*$, we may write a one-loop expression for the value
of the gauge couplings at the EW scale
\begin{eqnarray}
\nonumber
\alpha^{-1}_i(M_Z)&=&\alpha^{-1}_*(M_*)+\frac{1}{2\pi}\Bigg(
\alpha_i \ln\frac{m_{\rm SUSY}}{M_Z}
+\beta_i \ln\frac{M_*}{M_Z}\\
&+&\gamma_i \sum^{N_l}_{n=0}\ln\frac{M_*}{(2n+2)M_c}
+\delta_i \sum^{N_l}_{n=0}\ln\frac{M_*}{(2n+1)M_c}\Bigg),
\end{eqnarray}
where $(\alpha_1,\alpha_2,\alpha_3)=(-5/2,-25/6,-4)$,
$(\beta_1,\beta_2,\beta_3)=(33/5,1,-3)$ are usual MSSM coefficients
and
$(\gamma_1,\gamma_2,\gamma_3)=(4,-2,-5)$ and
$(\delta_1,\delta_2,\delta_3)=(-18,-12,-9)$
correspond to odd and even KK modes of Higgs and gauge fields in our
model.\footnote{In order to
calculate the $\beta$ functions, we have included the Higgs and
gauge hypermultiplets in the bulk along with the matter fields on
the visible brane. This is a rough approximation and we are
ignoring possible additional fields in the gauge symmetry and SUSY
breaking sectors and also fields in the flavor sector of the theory.}
All KK modes below $M_*$, where $(2N_l + 2)M_c\leq M_*$, are included in the
sum on $n$.
We assume that the model is $SO(10)$ symmetric beyond $M_*$
and also that a more fundamental theory can justify the
termination of the sums at $M_*$.

The gauge coupling unification in the conventional GUTs with only
the MSSM particle content below the GUT scale is well established
within the error-bars of the experimental values of the couplings
at the EW scale. However, in our model, in addition to the MSSM
states, we have a tower of heavy KK modes that are GUT
non-degenerate. Such KK modes alter the running of the couplings.
As a result the conventional 4-d GUT unification can not coexist
with the assumed unification in our model. In the following, we
assume that the gauge couplings exactly unify in our model and
based on that, we try to estimate the amount of the
non-unification of the couplings in the conventional GUTs. We show
that the amount of the non-unification is small enough to be
justified within the experimental error-bars of the couplings 
and our approximations. We
thus have a model that gives a consistent picture of the gauge
coupling unification.

Define $M_*=(2N_l+2) M_c$ to be the scale at which all three gauge
couplings unify. As an example, for $\alpha_1$ and $\alpha_2$ we can write
\begin{eqnarray}
\label{eq:alpha12a}
(\alpha^{-1}_1-\alpha^{-1}_2)(M_Z)=\frac{1}{2\pi}\Bigg(
\frac{5}{3}\ln\frac{m_{\rm SUSY}}{M_Z}
+\frac{28}{5}\ln\frac{(2N_l+2)M_c}{M_Z}
-6\sum^{N_l}_{n=0}\ln\frac{(2n+2)}{(2n+1)} \Bigg).
\end{eqnarray}
In order to compare this with the conventional 4-d GUTs, we define
$M_G$ to be the scale where $\alpha_1$ and $\alpha_2$ meet and
take the value $\alpha_G$. We assume that $\alpha_3$ does not
exactly unify with the other two couplings and we parametrize the
non-unification by a small parameter $\xi$ given by
$\alpha^{-1}_3(M_G)=\alpha^{-1}_G+\xi$. A similar formula to Eq.
(\ref{eq:alpha12a}) can be written in the conventional
4-dimensional GUTs:
\begin{eqnarray}
\label{eq:alpha12b}
(\alpha^{-1}_1-\alpha^{-1}_2)(M_Z)=\frac{1}{2\pi}\Bigg(
\frac{5}{3}\ln\frac{m_{\rm SUSY}}{M_Z}
+\frac{28}{5}\ln\frac{M_G}{M_Z}
\Bigg).
\end{eqnarray}
By comparing Eqs. (\ref{eq:alpha12a}) and (\ref{eq:alpha12b}) we
can calculate the value of the compactification scale:
\begin{eqnarray}
\label{eq:compa}
\ln \frac{M_c}{M_G}=
\frac{15}{14}\sum^{N_l}_{n=0}\ln\frac{(2n+2)}{(2n+1)}
-\ln(2N_l+2).
\end{eqnarray}
For example, if we take $N_l=4~(M_*=10M_c)$, for the range $1\times
10^{16}<M_G<3\times 10^{16}$ the compactification scale comes to
be in the range $4.5\times 10^{15}<M_c<1.3\times 10^{16}$.

In order to estimate $\xi$, we write equations similar to Eqs.
(\ref{eq:alpha12a}), (\ref{eq:alpha12b}) for $\alpha_1$ and
$\alpha_3$ and solve for the compactification scale.
\begin{eqnarray}
\label{eq:compb} \ln \frac{M_c}{M_G}=
\frac{45}{48}\sum^{N_l}_{n=0}\ln\frac{(2n+2)}{(2n+1)}
-\ln(2N_l+2)-\frac{5}{24}\pi\xi.
\end{eqnarray}
From Eqs. (\ref{eq:compa}),
(\ref{eq:compb}) one can calculate the level of non-unification
$\xi\approx -0.3$.\footnote{Note, $\xi$ is an increasing function of $N_l$, however 
this dependence is quite mild.} This is a small correction, less than $1.5 \%$
of the GUT scale value of the gauge coupling constant $\alpha^{-1}_G\approx 24$.
However, we note that our estimates are at one 
loop and subject to a few percent correction coming from higher 
loops. The fact that our estimated correction is not too large 
gives us hope that our model is consistent with the gauge coupling 
unification picture and the experimental values at the EW scale. 
In section \ref{sec:protondecay} we will show that 
large values of $N_l$ (larger than $\sim 30$) would be inconsistent with 
the current experimental limits on proton decay.

We note that the gauge group on the hidden brane is
only $SO(6)\times SO(4)$. In fact, one may write terms on
the hidden brane that do not respect the full $SO(10)$ symmetry 
of the bulk that is necessary for complete unification of the gauge 
couplings. Ref. \cite{Hall+Nomura} shows that the effects of 
these terms are small enough to be ignored in these models.

\section{Breaking the $\cal{N}$~=~1  supersymmetry}
\label{sec:susybreak}
In the previous sections, we benefitted from orbifolding to reduce the
amount of supersymmetry. However, we are still left with
$\cal{N}$~=~1 SUSY that survives the compactification. Since
there is yet no experimental evidence of SUSY particles at
current collider energies, the $\cal{N}$~=~1 SUSY must be broken
at scale of a TeV or higher. In this paper, we assume that SUSY is
broken by the vacuum expectation value of an $SO(6)\times SO(4)$
singlet chiral superfield $X$ that is localized on the hidden brane.
\beq
\langle X \rangle=\theta^2 F_X
\eeq
$X$ can couple directly to the gauge fields on the hidden brane through
the ultraviolet scale suppressed terms
\begin{eqnarray}
\nonumber
\lagrange_5&=&\frac{1}{2}\{\delta(x_5-\pi R/2)+\delta(x_5+\pi R/2)\}
\int{d^2\theta}\Big(\lambda_6^\prime\frac{X}{M^2_*}W^{i\alpha}
W^i_\alpha\\
&+&\lambda_4^\prime\frac{X}{M^2_*}W^{j\alpha}W^j_\alpha+h.c.\Big),
\end{eqnarray}
where index $i(j)$ runs over the number of gauge fields of the
$SO(6)~(SO(4))$
symmetry group.
This will give universal masses to the gauginos of $SO(6)$ and $SO(4)$
gauge groups separately,
\beq
\label{eq:gauginos}
M_6=\frac{\lambda_6^\prime F_X M_c}{M^2_*}, \qquad
M_4=\frac{\lambda_4^\prime F_X M_c}{M^2_*}.
\eeq
The factor $M_c$ is from the wave function normalization
of the 4-dimensional gaugino fields. The masses of the gauginos
of the MSSM $(M_1, M_2, M_3)$ are given as
\beq
M_1=\frac{2}{5}M_6+\frac{3}{5}M_4,\qquad M_2=M_4,\qquad M_3=M_6.
\eeq
The special form of $M_1$ is related to the fact that the hypercharge
operator is expressed as
$Y=\sqrt{\frac{2}{5}}(B-L)-\sqrt{\frac{3}{5}}t_{3R}$ where $B-L$ and
$t_{3R}$ are the generators of the $SO(6)$ and $SO(4)$ symmetry as
discussed in the appendix.
The couplings of $X$ to the fields on the visible brane are suppressed
at short distances by locality. The soft SUSY breaking scalar masses
and trilinear couplings are negligible at the GUT scale.
By RGE running down to the EW scale,
they receive large contributions from the gaugino mass terms and
acquire finite values. These contributions to the scalar masses are
flavor blind and thus do not cause large flavor changing neutral
currents. Such a scenario for mediating supersymmetry
breaking is called gaugino mediation \cite{gMSB}. 
Minimal gaugino mediation is characterized by
finite universal gaugino masses and negligible trilinear
couplings and scalar masses at the GUT scale.
Here, we have a special case of the non-universal gaugino
mediation where the Bino mass $M_1$
is completely constrained by the Wino mass $M_2$ and the gluino mass
$M_3$.
The usual
universal gagino mediation models predict the stau to be the lightest
supersymmetric particle (LSP). However stau is charged and is   
strongly disfavored by experimental data as the LSP. The   
partial non-universality of the gaugino masses in our model might 
provide a solution to this problem.

The SUSY breaking mechanism on the hidden brane can be easily linked to
the Higgs mechanism responsible for partial breaking of the gauge
symmetry on the visible brane. This is achieved via the shining mechanism
\cite{shining}. Consider a massive gauge singlet hypermultiplet in the
bulk
containing a pair of chiral superfields $\Phi$ and $\Phi^c$. The
hypermultiplet couples to the field $X$ on the hidden brane and to the
$16, \bar{16}$ on the visible brane. The action can be expressed as
\begin{eqnarray}
S&=&\int{d^5x}\Bigg(\int d^4\theta
[\Phi^\dagger\Phi+\Phi^{c\dagger}\Phi^c]
+\int d^2\theta [\Phi^c(m+\partial_5)\Phi \\
\nonumber
&+&\frac{1}{2}(\delta(x_5)+\delta(x_5-\pi R))16~\bar{16}\Phi^c
+\frac{1}{2}(\delta(x_5-\pi R/2)+\delta(x_5+\pi R/2))X\Phi]\Bigg).
\end{eqnarray}
$\langle 16~\bar{16}\rangle$ acts as a source and $\Phi$
develops a non-trivial profile in the bulk. The shining
mechanism gives a SUSY breaking F-term vev to the field
$X$ on the hidden brane,
\begin{equation}
F_X\approx \langle 16~\bar{16}\rangle \exp({-\frac{\pi m R}{2}}).
\end{equation}
Note that the appropriate phenomenology with TeV scale SUSY masses
in Eq. (\ref{eq:gauginos}) is possible with
$\sqrt{F_X}\approx 10^{11}-10^{12}$ GeV. This is a
mass scale that is almost $10^5$ times smaller than the GUT scale. The
SUSY
breaking scale seems to have a completely independent nature from the
grand unification scale. However, the power of shining mechanism is that
the
SUSY breaking scale is generated from the compactification scale and 
the undesirable large hierarchy between the compactification and 
SUSY scales is cured by taking $m\sim 10M_c$ which is close to its natural 
value, i.e. the ultraviolet scale $\sim M_*$. We emphasize 
that in our model, in addition
to the ultraviolet scale, we have only the compactification scale
and the SUSY breaking is dynamically generated via the shining mechanism at
the appropriate scale. As a result, the GUT breaking and SUSY breaking have
common origins in our model and are linked together. 
\section{Generating the $\mu$ term, solving the $\mu$ problem}
\label{sec:muterm}
The $\mu$ term in the superpotential of the MSSM couples the two light
Higgs doublet superfields with a mass dimension one parameter, $\mu$.
Since the $\mu$ term respects supersymmetry and the gauge symmetries
of the MSSM, one expects it to be of order the ultraviolet scale
in the theory. However, for various phenomenological
reasons, $\mu$ parameter needs to be of order the EW scale.
The difficulty in generating the $\mu$ parameter at the right
scale is called the $\mu$ problem. There is another parameter
in the MSSM ($B_\mu$) with mass dimension two that couples the
Higgs doublets in the soft SUSY breaking Lagrangian. $B_\mu$
is also expected to be of order the EW scale, squared. Our
goal is to set up our model in order to have both $\mu$ 
and $B_\mu$ parameters at the right scale.

It was noted earlier in this paper that by assigning PQ charges to
the 10 dimensional Higgs multiplets in the bulk, we avoided terms like
$M_c 10_H 10_H$ on the branes that potentially give rise to a very
large $\mu$ parameter. Our model presents a solution to the $\mu$ problem
while keeping $B_\mu$ under control.
One can imagine an $SO(6)\times SO(4)$ singlet chiral superfield $Y$ on
the hidden brane with PQ charge $Q_Y=+2$. An ultraviolet suppressed
term of the form \cite{AMmu}
\begin{eqnarray}
\label{eq:muterm}
\int{d^5x}\frac{1}{2}\{\delta(x_5-\pi
R/2)+\delta(x_5+\pi R/2)\}
\int{d^4\theta}(\frac{Y^\dagger}{M^2_*}
4_H 4_H)+h.c.
\end{eqnarray}
is allowed on the hidden brane and can result in a $\mu$ term
\beq
\mu\sim \frac{F_Y M_c}{M^2_*}
\eeq
if
\beq
\langle Y \rangle=\theta^2 F_Y.
\eeq
Comparing this with the result in Eq. (\ref{eq:gauginos}), one notices that
the value of $\mu$ is at the right scale provided that the SUSY breaking
vevs of $Y$ and $X$ are comparable.\footnote{We proposed that $X$ gets a   
SUSY breaking vev through the shining mechanism. However, we assume that 
there exists a potential on the hidden brane for $X$ and $Y$ that 
relates the vev of $Y$ to the vev of $X$ at the same scale.}
It is important to note that
$B_\mu$ is also generated at this order since
\beq
\label{eq:Bmuterm}
\int{d^4\theta}\frac{X Y^\dagger}{M^3_*}4_H 4_H+h.c.
\eeq
is allowed by the PQ symmetry ($Q_X = 0$). 
\beq
B_\mu\sim \frac{F^2_Y M_c}{M^3_*}.
\eeq
It is now evident from $\sqrt{B_\mu}/\mu\sim \sqrt{M_*/M_c}\sim 3$ that
$B_\mu$ is also generated at the right scale.

\section{Proton decay}
\label{sec:protondecay}
Proton decay in grand unified theories can happen through dimension 
six operators coming from the exchange of the X gauge bosons. The mass 
of the $X$ gauge bosons must be large enough to bring the predicted 
proton decay rate below the current experimental bounds.
Note that the mass of the X gauge bosons in our model is the compactification
scale and so the experimental bounds on the proton decay set limits on the
compactification scale. 

Super-Kamiokande puts a bound on 
$\tau_{p\to e^+\pi^0} > 4.4\times 10^{33}$ ($90\%$ CL) \cite{SuperK} 
and this translates into a
limit on the compactification scale $M_c>6\times 10^{15}$ GeV.
In our model, the size of the compactification scale from Eq. (\ref{eq:compa}) 
is a decreasing function of $N_l$. 
Therefore, the current experimental bound on $p\to e^+\pi^0$
gives an upper bound on $N_l \sim 30$. However, it should be noted that 
this is a quite conservative limit. Improving the experimental bounds will 
result in a lower upper limit on the value of $N_l$.
Thus the predictions of our model can be tested in the current or the next 
generation proton decay experiments.

There are other possible dangerous sources for proton decay in the
$SO(10)$ grand unified theories. Large contributions to proton decay
are expected from the Higgs triplet exchanges. 
We show that unlike the conventional $SO(10)$ GUTs, the Higgs
triplet contribution vanishes in our 5-dimensional model.\footnote{The 
discussion of proton decay in our $SO(10)$ model is 
similar to the case of $SU(5)$ discussed in Ref. \cite{Hall+Nomura}.}
Consider the relevant terms coming from Eq. (\ref{eq:matterb})
for the proton decay through the Higgs triplets
\begin{eqnarray}
\label{eq:pdecaymatter}
\frac{1}{2} Q Q t^{(2n+1)}_H+ U E t^{(2n+1)}_H+
Q L \bar{t}^{(2n+1)}_H+ U D \bar{t}^{(2n+1)}_H.
\end{eqnarray}
The only mass terms consistent with the PQ symmetry in our model 
are coming from the bulk contribution
Eq. (\ref{eq:action}) of the form
\beq
M_c(t^{\prime(2n+1)}_H \bar{t}^{(2n+1)}_H+t^{(2n+1)}_H
\bar{t}^{\prime(2n+1)}_H).
\eeq
After integrating out the GUT-scale heavy fields  $t_H, \bar{t}_H,
t^\prime_H, \bar{t}^\prime_H$ we observe that we get no 
terms of the type $QQQL$ or
$\bar{U}\bar{D}\bar{U}\bar{E}$ and thus no proton decay
from dimension five operators. 
It is worth to note that 
even an additional mass term for the Higgs triplets of the form 
\beq
M_c t^{(2n+1)}_H \bar{t}^{(2n+1)}_H
\eeq
does not affect the proton decay result that we just mentioned.
Another possible source for proton decay is the operator 
$16~16~16~16$ on the visible brane where $16$s  represent generations 
of matter fields. This term is also forbidden by the PQ symmetry.

The conventional SUSY GUTs are pushed to their limits by the updated
experimental bounds on the proton decay. We showed that our model 
survives the current experimental bounds yet may be tested in the
near future.

\section{Conclusions}
\label{sec:conclusions}
We have constructed an $SO(10)$ supersymmetric grand
unified model in 5 dimesions. A gauge supermultiplet
and a single Higgs hypermultiplet live in the 5-dimensional
bulk. The extra dimension is compactified on an orbifold
$S^1/(Z_2 \times Z_2^\prime)$ which has two fixed points:
the visible and the hidden branes. The matter supermultiplets   
are confined on the visible brane. By orbifolding $\cal{N}$~=~2
supersymmetry is reduced to
$\cal{N}$~=~1, and the $SO(10)$ gauge symmetry is reduced
to $SO(6)\times SO(4)$. Unfortunately it is not
possible to reduce $SO(10)$ down to the SM gauge symmetry by an abelian
orbifolding and the further reduction
of the gauge symmetry to that of SM is achieved by ordinary
Higgs mechanism.
However, we argue that this does not have to be viewed as a downside
of the model.
The Higgs mechanism responsible for partial breaking of the
gauge symmetry
can be used to trigger the breaking of the remaining $\cal{N}$~=~1
supersymmetry via the shining mechanism. The GUT breaking vev of the
$16, \bar{16}$ pair living on the visible brane
acts as a source for the shining mechanism which gives a SUSY breaking
F-term vev to the field X living on the hidden brane.
The large hierarchy between the SUSY breaking scale
and the GUT breaking scale is achieved by the exponential
suppression of the SUSY breaking vev with respect to the source.
The F-term vev of $X$ gives soft SUSY breaking masses to gauginos.
The breaking of supersymmetry is communicated to squarks and
sleptons through gaugino mediation and so the flavor changing neutral
currents are suppressed. The specific feature of this model is the
non-universality of gauguino masses.
The Bino mass $M_1$
is completely constrained by the Wino mass $M_2$ and the gluino mass
$M_3$ which is  
dictated by the $SO(6)\times SO(4)$ gauge symmetry on the hidden brane.
This might provide a solution to the usual problem of universal gaugino
mediation models which typically predict stau to be the LSP.
The $\mu$ and $B_\mu$ terms are also generated on the hidden brane 
at the scale
that can also  be linked to the GUT symmetry breaking vev.
Operators of dimenion 5 which could lead to proton decay are forbidden 
by PQ
symmetry. Proton decay through dimension 6 operators is enhanced 
compared to
conventional GUTs and can be seen in current or next generation proton 
decay experiments.

\vspace{.5cm}

\noindent {\bf Acknowledgements:} We would like to thank S. Raby 
and T. Bla\v zek for discussions. R.D. thanks the Physics Department 
at the University of Bonn and the Theory Division at CERN for their 
kind hospitality while working on this project. R.D. is partially 
supported by DOE grant DOE/ER/01545-818. A.M. would like to thank
the organizers of SUSY and Extra Dimensions 2001 and The ANL HEP 
Theory Group where part of this work was done. A.M. is supported 
in part by the National Science Foundation under Grant PHY-0071054.

\section*{Appendix}
\label{sec:appendix} In order to better
understand the breaking of the gauge group and the notation used
in this paper, we present a short review of $SO(10)$ group theory
in this appendix. A generator of $SO(10)$ group is an imaginary
$10\times 10$ antisymmetric matrix and may be written as
\begin{eqnarray}
\label{eq:so10generator}
t=A_{5\times 5}\otimes 1_{2\times 2}+B_{5\times 5}\otimes \sigma^2
+C_{5\times 5}\otimes \sigma^1+D_{5\times 5}\otimes \sigma^3,
\end{eqnarray}
where $\sigma^a$ are the Pauli matrices. $A_{5\times 5}$, $C_{5\times 5}$
and $D_{5\times 5}$ are $5\times 5$ imaginary antisymmetric matrices while
$B_{5\times 5}$ is a $5\times 5$ real symmetric matrix. It is possible to
classify the generators of $SO(10)$ algebra based on whether they
commute $(t_+)$ or anti-commute $(t_-)$ with
$P^\prime= \, \rm diag \, (-1,-1,-1,+1,+1)\otimes 1_{2\times 2}$.
The $SO(10)$ generators that commute with $P^\prime$ can always
be written as
\begin{eqnarray}
\label{eq:so10plus}
\nonumber
t_+&=&\left(\begin{array}{cc} A_{3\times 3}&0\\
0&A_{2\times 2}\end{array}\right)\otimes 1_{2\times 2}+
\left(\begin{array}{cc} B_{3\times 3}&0\\
0&B_{2\times 2}\end{array}\right)\otimes \sigma^2\\
&+&\left(\begin{array}{cc} C_{3\times 3}&0\\
0&C_{2\times 2}\end{array}\right)\otimes \sigma^1+
\left(\begin{array}{cc} D_{3\times 3}&0\\
0&D_{2\times 2}\end{array}\right)\otimes \sigma^3,
\end{eqnarray}
where $A_{3\times 3}, C_{3\times 3}, D_{3\times 3}
(A_{2\times 2}, C_{2\times 2}, D_{2\times 2})$ are
$3\times 3 (2\times 2)$ imaginary antisymmetric matrices and
$B_{3\times 3} (B_{2\times 2})$ is a $3\times 3 (2\times 2)$
real symmetric matrix. The rest of the $SO(10)$ generators
in Eq. (\ref{eq:so10generator}) anticommute with $P^\prime$.

The $t_+$ generators of $SO(10)$ in Eq. (\ref{eq:so10plus})
are in fact the elements of the $SO(6)\times SO(4)$
subalgebra of $SO(10)$. Eq. (\ref{eq:so10generator}) is a representation
of the $SO(10)$ generators based on $5\times 5$ matrices times Pauli
sigma matrices. In the same way, Eq. (\ref{eq:so10plus}) represents the
$SO(6)~(SO(4))$ generators based on $3\times 3 \, (2\times 2)$ matrices
times Pauli sigma matrices. Note, that $SO(6)\sim
SU(4)$
and $SO(4)\sim SU(2)_L\times SU(2)_R$. In fact, $SU(4)$ contains
$SU(3)\times U(1)_{B-L}$ where $SU(3)$ is the SM QCD gauge group and
$U(1)_{B-L}$ is associated with the baryon number minus lepton number
generator.

The $SU(3)\times U(1)_{B-L}$ subalgebra consists of $t_+$ generators
in Eq. (\ref{eq:so10plus})
with all entries zero except $A_{3\times 3}$ and $B_{3\times 3}$. In fact,
$U(1)_{B-L}$ is generated by $A_{3\times 3}=0$ and
$B_{3\times 3}=1_{3\times 3}$.
$SU(2)_L\times SU(2)_R$ contains the  gauge group
$SU(2)_L\times U(1)_{t_{3R}}$. In our notation, $U(1)_{t_{3R}}$
is generated by $A_{2\times 2}=0$ and $B_{2\times 2}=1_{2\times 2}$.
Hypercharge symmetry is a combination of $U(1)_{B-L}$ and
$U(1)_{t_{3R}}$ gauge symmetries. Note that one needs to define a
proper normalization of the $SO(10)$ generators. In this paper we
use the usual convention $Tr[t^at^b]=\delta^{ab}/2$. The properly
normalized $B-L=\frac{1}{\sqrt{12}} \, \rm diag \, (1,1,1,0,0)\otimes
\sigma^2$
and $t_{3R}=\frac{1}{\sqrt{8}} \, \rm diag \, (0,0,0,1,1)\otimes \sigma^2$
combine into the hypercharge generator
$Y=\sqrt{\frac{3}{40}} \, \rm diag \, (2/3,2/3,2/3,-1,-1)\otimes \sigma^2$
where $Y=\sqrt{\frac{2}{5}}(B-L)-\sqrt{\frac{3}{5}}t_{3R}$.

The 10-dimensional Higgs representation of $SO(10)$ is realized under
$SO(6)\times SO(4)$ as $10=6\oplus 4$. Under the standard
model gauge group however, we have $6=t\oplus \bar{t}$ and
$4=d\oplus \bar{d}$ where $t~(\bar{t})$ is a
color triplet~(anti-triplet) and $d, \bar{d}$ are weak
doublets. It is essential to note that under
$P^\prime= \, \rm diag \, (-1,-1,-1,+1,+1)\otimes 1_{2\times 2}$
doublets $d$ and
$\bar{d}$ have positive parities while triplets $t$ and $\bar{t}$
have negative parities. This is in fact the essence
of the doublet-triplet splitting mechanism used in this paper.

\newpage

% ---- Bibliography ----
%

\end{document}